\documentstyle[11pt,aaspp4]{article}

\def\degrees{$^\circ$}

\def\spose#1{\hbox to 0pt{#1\hss}}
     
\def\={\overline}

\newcount\notenumber
\newcount\eqnumber
\newcount\fignumber
\newbox\abstr
\newbox\figca

\def\etal{{\it et al. }}
\def\eg{{\it e.g., }}
\def\ie{{\it i.e., }}
\def\cf{{\it cf. }}

\def\note#1{\footnote{$^{\the\notenumber}$}{#1}\global\advance\notenumber by 1}
\def\foot#1{\raise3pt\hbox{\eightrm \the\notenumber}
     \hfil\par\vskip3pt\hrule\vskip6pt
     \noindent\raise3pt\hbox{\eightrm \the\notenumber}
     #1\par\vskip6pt\hrule\vskip3pt\noindent\global\advance\notenumber by 1}

\def\Dt{\spose{\raise 1.5ex\hbox{\hskip3pt$\mathchar"201$}}}    
\def\dt{\spose{\raise 1.0ex\hbox{\hskip2pt$\mathchar"201$}}}    

\def\new{{\rm\chaphead\the\eqnumber}\global\advance\eqnumber by 1}
\def\ref#1{\advance\eqnumber by -#1 \chaphead\the\eqnumber
     \advance\eqnumber by #1 }
\def\last{\advance\eqnumber by -1 {\rm\chaphead\the\eqnumber}\advance
     \eqnumber by 1}
\def\eqnam#1{\xdef#1{\chaphead\the\eqnumber}}
     
\def\nfig{\chaphead\the\fignumber\global\advance\fignumber by 1}
\def\nfiga#1{\chaphead\the\fignumber{#1}\global\advance\fignumber by 1}
\def\rfig#1{\advance\fignumber by -#1 \chaphead\the\fignumber
     \advance\fignumber by #1}
\def\refindent{\par\noindent\parskip=4pt\hangindent=3pc\hangafter=1 }

\def\apj#1#2#3{\refindent#1,  {ApJ,\ }{#2}, #3}
\def\apjsup#1#2#3{\refindent#1,  {ApJS\ }{#2}, #3}

\def\aas#1#2#3{\refindent#1,  { Bull. Am. Astr. Soc.,\ }{#2}, #3}

\def\mn#1#2#3{\refindent#1,  { MNRAS,\ }{#2}, #3}

\def\annrev#1#2#3{\refindent#1, { ARA \& A,\ }
{\bf2}, #3}
\def\aj#1#2#3{\refindent#1,  { AJ,\  }{#2}, #3}

\def\aa#1#2#3{\refindent#1,  { AA,\ }{#2}, #3}

\def\pasp#1#2#3{\refindent#1,  { PASP,\ }{#2}, #3}

\def\refbook#1{\refindent#1}

\def\ltsim{\mathrel{\spose{\lower 3pt\hbox{$\mathchar"218$}}
     \raise 2.0pt\hbox{$\mathchar"13C$}}}
\def\gtsim{\mathrel{\spose{\lower 3pt\hbox{$\mathchar"218$}}
     \raise 2.0pt\hbox{$\mathchar"13E$}}}
     
\def\sec{\hbox{$^s$\hskip-3pt .}}

\def\apequal{\mathrel{\spose{\lower 1pt\hbox{$\mathchar"218$}}
     \raise 2.0pt\hbox{$\mathchar"218$}}}
\newbox\grsign \setbox\grsign=\hbox{$>$} \newdimen\grdimen \grdimen=\ht\grsign
\newbox\simlessbox \newbox\simgreatbox
\setbox\simgreatbox=\hbox{\raise.5ex\hbox{$>$}\llap
     {\lower.5ex\hbox{$\sim$}}}\ht1=\grdimen\dp1=0pt
\setbox\simlessbox=\hbox{\raise.5ex\hbox{$<$}\llap
     {\lower.5ex\hbox{$\sim$}}}\ht2=\grdimen\dp2=0pt

\def\Kp{K$^\prime$\ }
\def\etal{{\it et al.\ }}

\begin{document}

\def\Kp{K$^\prime$\ }

\def\etal{{\it et al.\ }}

\title{\bf A Digital Photometric Survey of the Magellanic Clouds: \break
First Results From One Million Stars$^1$}

\author{Dennis Zaritsky\altaffilmark{2}, Jason Harris\altaffilmark{2},
and Ian Thompson\altaffilmark{3}}

\bigskip

\affil{$^2$ UCO/Lick Observatory and Board of Astronomy and Astrophysics,
Univ. of California at Santa Cruz, Santa Cruz, CA 95064. Email:
dennis@ucolick.org, jharris@ucolick.org}

\medskip

\affil{$^3$ Carnegie Observatories, 813 Santa Barbara St., Pasadena CA 91101,
Email: ian@ociw.edu}

\vskip 1in

$^1$ Lick Bulletin No. 1363

\vskip 1in

\setcounter{footnote}{4}

\begin{abstract}
We present the first results from, and a complete description of, our
ongoing $UBVI$ digital photometric survey of the Magellanic 
Clouds. In particular, we discuss the photometric quality
and automated reduction of a CCD survey
(magnitude limits, completeness, and astrometric accuracy)
that covers the central $8^\circ \times 8^\circ$ of
the Large Magellanic Cloud (LMC) and $4^\circ \times 4^\circ$ of the
Small Magellanic Cloud (SMC).
We discuss photometry of over 1 million
stars from the initial survey observations
(an area northwest of the LMC bar
covering $\sim 2^\circ \times 1.5^\circ$) and present a
deep stellar cluster catalog that contains about 45\% more
clusters than previously identified within this region.
Of the 68 clusters found, only 12 are also identified
as concentrations of ``old'', red clump stars. 
Furthermore, only three clusters
are identified solely on the basis of 
a concentration of red clump stars,
rather than as a concentration of luminous ($V < 21$) main sequence
stars. 
Extrapolating from the current data, we expect to obtain 
$B$ and $V$ photometry for 25 million
stars, and $U$ and $I$ photometry for 10 and 20 million stars,
respectively, over the entire survey area. 
\end{abstract}
\vfill\eject
\section{Introduction}

Our knowledge of galaxies and stellar
populations rests in large part on our understanding of the Large and Small
Magellanic Clouds (LMC and SMC). Due to their proximity (49 and 57
kpc respectively, Feast \& Walker 1987), 
relatively low line-of-sight extinction
(Schwering and Israel 1991; Oestreicher \& Schmidt-Kaler 1996),
range of stellar populations (cf. Olszewski, Suntzeff, \& Mateo 1996),
various interstellar gaseous environments, and 
difference in mean metallicity relative to each other and to the Milky 
Way (Lequeux \etal 1979), these two galaxies are our most promising targets for detailed studies
of galaxy evolution. Although the Clouds' proximity allows us to resolve 
individual stars well down the luminosity function even 
in ground-based images,
the Clouds have such a large angular extent on the sky that they are
difficult to observe in their entirety. 
Investigations are generally 
limited to small, well-chosen regions, to low
spatial resolution, or to shallow magnitude limits.
For example, even the ambitious photographic
surveys of the Clouds (\cf Hodge \& Wright 1967; Hatzidimitriou, 
Hawkins \& Gyldenkerne 1989), 
are either not very sensitive (Hodge \& Wright's plates
have an average limiting $V$ magnitude of 17) or lack resolution
(Gardiner \& Hatzidimitriou's (1992) 
COSMOS digitized scans of the ESO/SERC plates reach
$B = 21$, but the authors confine their analysis to an area outside
the SMC's main body,
the central 3\degrees\ by 4\degrees, due to crowding).
Furthermore, these and other large-area surveys, including
the ongoing MACHO survey (Alcock \etal 1996), are conducted with only
two optical filters,
which limits the baseline for extinction measurements 
and the study of the full range
of stellar spectral types. A
photometric survey of the Magellanic Clouds that spans the full 
optical wavelength range will enable us to better
understand our nearest galactic neighbors and provide a stepping-stone
for future, focused investigations.

The development of large CCD arrays and
a drift-scan camera (Zaritsky, Shectman, 
\& Bredthauer 1996) that can scan at the extreme
southerly declination of the Clouds ($\sim -$70\degrees) 
enables us to conduct efficiently a large-area, high-resolution, digital, 
multi-color photometric survey of the Clouds.
Our goal is to
image the central 8\degrees\ by 8\degrees\ of the LMC and
4\degrees\ by 4\degrees\ of the SMC in the $UBVI$ bandpasses. 
We report on our initial
observations, data reduction, and data analysis of a smaller ($\sim$
2\degrees\ by 1.5\degrees) area of the LMC, and discuss
the photometric and astrometric quality of the data in \S2. In \S3,
we illustrate
the usefulness of digital data by automatically identifying 
stellar clusters in the LMC from a stellar density image.
Using a simple unsharp masking technique, we increase the number of known
clusters in this region by about 45\%. Other results obtained from these
data will be reported elsewhere: for example,
the reddening properties and constraints on the distribution of dust
are described by 
Harris, Zaritsky, \& Thompson (1997).

\section{Observations and Data Reduction}

The observations for
this survey are to be obtained from a multiyear effort based at the
Las Campanas 1-m Swope telescope using the Great Circle
Camera (GCC; Zaritsky, Shectman, \& Bredthauer 1996) and
a thinned 2048$\times$2048 CCD. The GCC
enables us to drift scan at the declination of the Magellanic
Clouds with minimum image distortion by rotating and translating
a stage onto which the CCD dewar is mounted. The telescope remains
parked during an exposure,
but the CCD moves in such a way that a scan is performed along
a great circle on the sky, rather than on a polar circle of 
constant declination. Implementing the motion requires an accurate
rotational alignment of the camera and a precise measurement 
of the plate scale
(both in terms of pixels per arcsec and in terms of stage movement per
arcsec).
Details of the scan geometry and camera are discussed by Zaritsky,
Shectman, and Bredthauer.
The effective exposure
time is fixed by the time required for the sky to drift across
the field-of-view of the stationary telescope (about
240 sec at the declination of the Clouds).
Photometric standard stars are observed in a similar manner, which ensures that
they have a well defined exposure time (the GCC does not
have a precise shutter).

The two
Clouds are divided into scan-sized sections,
where each scan is about 2$^\circ$ long and 24 arcmin wide. 
Such scans take about 25 min to complete.
The 8\degrees\ by 8\degrees\ LMC area is divided into 4 such scans in RA
and 23 in Dec. The scans 
overlap by about 30 arcsec in right ascension and
declination, which enables us to 
tie together the photometric and astrometric solutions. The 
CCD has a 0.7 arcsec pixel$^{-1}$ scale and the typical seeing at the
telescope is between 1.2 and 1.8 arcsec. When the survey progresses and the
seeing statistics are better determined, we will select an upper limit
on the seeing for acceptable data. Scan quality will be judged after
reduction and unsatisfactory scans will be reobserved in subsequent
observing runs.

In this paper, we present results from the LMC scans numbered
58, 62, 66, and 70, for which
the start (westernmost) coordinates are listed in Table 1. 
These data were obtained
during Nov 13-23, 1995. The box in Figure 1 outlines 
the area observed and numbers the scans, superposed
on a greyscale image of the central portion of the LMC. 
When complete, the survey
will cover a larger area of the LMC than that covered by the Figure.
The data consist of Johnson $U$, Harris $B$, $V$ and Johnson $I$ exposures
taken mostly in photometric conditions. Scans taken in moderately
non-photometric
conditions (\eg slight cirrus)
are corrected using the overlap regions with photometric scans.
Details of the photometric matching of scans are presented
in \S2.1. The region observed
was selected to provide a compromise in the 
level of stellar crowding and to avoid
regions of high nebulosity.
Regions with high stellar density and nebulosity will be observed
as part of the complete survey. 

\subsection{Reduction Procedure}

The quantity of data generated in this survey
mandates that we automate the reduction.
We constructed an algorithm that subtracts the bias using
the overscan columns, ``flattens'' the image (along columns) using
a drift scan of the twilight sky, fixes bad columns (which are defined by the
user and are stable during a run) via interpolation, and divides each
scan into twenty manageable sections ($\sim 1000 \times 1000$ pixels) 
that overlap each other by about 50 pixels.
We refer to the small working
images as subscans and each scan is divided into a 10 
by 2 array of subscans. The photometry is obtained using DAOPHOT II
(cf. Stetson 1987 for a description of the original DAOPHOT package;
the version we use dates from April 1991)
and coordinate solutions are derived using the
Magellanic Catalogue of Stars (MACS; Tucholke, de Boer, \& Seitter
1996). The photometric catalogs constructed from different subscans, and then
from different scans, are pasted together
to produce a final, uniform
catalog. Details of this procedure are discussed next.

Crowded field stellar photometry is nearly an art form, but has 
become relatively standardized in great part due to the availability
of the DAOPHOT suite of data analysis routines. We iteratively
selected values for the appropriate parameters
and developed an algorithm, using the DAOPHOT II routines,
that minimizes the residuals in the star-subtracted image. 
Basically, we iterate three times through the reduction of every subscan 
to find a suitable point-spread function (PSF) and measure
PSF photometry. The first pass is done using only a subset of
the brightest stars
(V $\ltsim$ 18.5) and an
analytic Moffat function as the PSF model (a 2-D elliptical with
an unspecified position angle). The stars used to derive the 
PSF are chosen from among
the brightest non-saturated stars and then ranked in order of
contamination by nearby neighbors. We use up to 100 relatively isolated stars
that are at least 30 pixels from the image edge, 
within 4.5 magnitudes of the brightest non-saturated
star (effectively, $12 \ltsim V \ltsim 17$), and
in regions where the mean background sky is less than 1$\sigma$ larger than
the mean sky in the image (to avoid regions with strong nebulosity).
Once the
sample of PSF stars is defined, it is used through the three iterations.
In each iteration, nearby neighboring stars identified in the previous
pass are subtracted before the PSF is calculated.
The second pass is done using the same PSF model,
but adding an empirical lookup table for deviations
from the analytic form.
The final pass is done with a spatially quadratically
variable Moffat function plus an analytic lookup table. 

The PSF can vary along the length of a scan, either because the GCC motion is
not entirely compensating for field motion (as can happen if the
stage alignment is not quite perfect), or because the seeing
changes with time. 
However, we find that the PSF in sections of about 1000 pixels square
is well-behaved (\eg it can be modeled with a spatially quadratically variable 
function without leaving significant positional dependent residuals in the
star subtracted image). This finding and the issue of 
image manageability 
dictate that the reduction be done on subscans of roughly 1000 pixels
on a side. We present data from the first (westward) 9/10ths of each
scan because the scans are systematically poor at the scan end (the
last 1/10th of the scan).
This ``tracking'' problem has since been corrected, but
we exclude the poor quality data from further discussion.

The result of executing the above procedure is a list of $x,y$ positions and 
instrumental magnitudes that are based on the amplitude of the
best-fit PSF
for each star. Aperture corrections, which 
convert that magnitude into an instrumental total magnitude,
are computed using images from which all stars, except
those used to derive the PSF, are subtracted. The aperture correction
is calculated by evaluating the magnitude of every PSF star
in concentric apertures of increasing radius and extrapolating 
the magnitude vs. aperture size relationship to infinite radius.
The correction is applied to
all stellar magnitudes from the relevant subscan. Any errors in
aperture corrections will become evident when comparing photometry
from overlapping subscans.

To match the photometry for stars in one subscan 
either to those of adjacent subscans in the same filter or to overlapping
subscans in 
other filters, we need to convert the stellar $x,y$ positions
into equatorial coordinates. Subscans and scans cannot simply be
registered
and combined because the geometric transformation between pixel
and equatorial coordinates depends on the motion of the GCC and
can be slightly different even for scans of the same region. For
example, the plate scale, and therefore the CCD readout rate, 
changes among filters.
To convert between $x,y$ positions and equatorial coordinates,
we utilize the
Magellanic Catalogue of Stars (MACS; Tucholke, de Boer, \& Seitter
1996). We minimize the residuals between the coordinates of
astrometric standards from the MACS and those derived for the same stars
from a gnomic-projection of the equatorial coordinates onto our $x,y$-space.
We typically match between 20 and 40 MACS stars per subscan from which
we define the plate solution.
This procedure enables us to derive the scale,
orientation, and central coordinates of each subscan. We use these
solutions to convert the positions of all stars to RA and Dec.
The {\it rms} differences among the cataloged and measured positions
from our $V$ frames is typically $\ltsim$ 0.3 to 0.4
arcsec (\cf \S2.3.2 for more details). 

Using the equatorial coordinates, we combine
the stellar lists from the matching subscans to produce a list that
includes coordinates and magnitudes for the stars
in all four filters.
The $V$ data are adopted as 
the positional reference. Data from other
filters are matched to the $V$ data.
We match stars in two different filters
by using a search radius
equal to twice the {\it rms} positional scatter 
of the worst astrometric solution for the two relevant subscans.
The nearest star in the frame being matched to each
star in the $V$ frame that is within the search radius
is adopted as the matching star. 
After completing the first attempt at matching all of the stars
between the two subscans, 
we apply a positional offset to the data from the 
second frame (no scale change or rotation is allowed)
so that the mean positions of the matched stars in the second subscan
agree with those of the matched stars in the
$V$ subscan.
Next, we increase the search radius by 0\sec1 and redo the search.
We continue incrementing the search radius by 0\sec1 until the
former iteration produced at least 99\% of the matches found in the
latter iteration. We repeat this procedure with the
photometry from each filter relative to $V$ data.
Unmatched stars are retained in
the catalog at this point. The scans in different filters
do not overlap precisely, so some
stars identified near the image edge in one filter 
may lie on a different scan or subscan.
Retaining stars that are found in only one filter enables us
to recover their matching stars when different subscans are matched.
This procedure results in lists of equatorial coordinates and
four-filter
photometry corresponding to each subscan.

The data from each subscan are combined with those from the 
adjacent subscans in RA
by iterating using the same matching algorithm. In this case,
we begin with a matching radius of 0\sec4 and increment
the search radius by 0\sec1 until subsequent searches identify
the same number of matches to within 1\%. If the mean positional
offset between subscans
is greater than one-half of the standard deviation of the stellar
positions, we recenter the subscans relative to each other and
redo the matching. We average the magnitudes
for matched stars. These are not two independent
measurements
of the magnitudes, because they are drawn from the same image, but the
measurements do
differ slightly due to the different derived PSFs and aperture
corrections. 
No mean magnitude offsets between subscans
are applied (the corrections were found to be too unstable and caused the
photometry along a scan to random walk). 
We repeat this procedure until each of the two sets of
nine subscans, for each scan, 
are matched in RA. We then match the two sets in Dec
following the same procedure. As mentioned above, 
because we retain stars that are
observed in at least one filter, we do not lose stars because of
slight misalignments between scans.
Once all of the subscans from one scan
are combined, then
the photometric calibration (see below)
is applied to the data from that scan. Finally, we
only retain stars for which at least $B$ and $V$ measurements 
are available in order to minimize spurious detections.

The data from the various full 
scans are matched to each other only after the photometric
calibration so that all scans are on the same photometric system when
combined. The stars in overlapping
scans are matched using the same algorithm that was used to match
stars in overlapping subscans, except that
we now apply a magnitude correction between scans because different
scans may have slight zero-point magnitude differences. Several
hundred (for the $U$ frames) to several thousand (for the $B,V,I$
frames) stars are typically
matched in the overlap
regions between scans. Data for each filter are matched independently. In the 
simplest case, a single constant magnitude offset between scans would
suffice. However, we find that there are slight, but systematic,
variations along scans. If severe ($\gtsim 0.1$ mag),
these fluctuations are possibly an indication
of nonphotometric conditions; if slight, these are possibly due to
changes in the PSF and the related aperture correction. In either
case, we need to map and remove the relative variations between scans
using a fiducial scan that is thought to be free of temporal magnitude
variations. We choose to reference all of the scans to
scan 62, which has high-quality images taken
in photometric conditions for all filters 
and includes an OB association with previously published photometry
for comparison. 

The procedure for mapping the magnitude variations between scans is
described next. We
fit a 5th order (for the $U$ frames) or
10th order polynomial to the magnitude differences between
matched stars versus RA
(only using magnitude differences that are within $5\sigma$ of
the mean difference). The fitted polynomial describes the mean
photometric
offset between the two scans as a function of RA.
The typical corrections
have small spatial variation and zero-point photometric corrections
$<$ 0.1 mag (cf. Figure 2). One particular scan (70 in $I$),
taken during nonphotometric conditions, shows fairly large
photometric variations along the scan.
The polynomial
fits are used to reference the photometry from the
three other scans to that from scan 62. For a small section of
scan 62, we can
check our photometry with published photometry 
(see discussion in 
\S2.4). Eventually,
our observations will extend over an area that is roughly
25 times the current area and will
overlap with a variety of published photometry.
Comparison with those data will confirm the overall photometry and 
whether the scans were properly referenced to each other.

\subsection{Standards}

The photometry is calibrated to the Landolt system (Landolt 1992). 
The details of the standard star
reductions, repeatability, and color terms are discussed below.
The photometric calibration is fairly standard and done within
DAOPHOT. We solve for $U-B$, $B-V$, $B-V$, and $V-I$ color terms for
$U$, $B$, $V$, and $I$, respectively, and for extinction as a function
of airmass, $Z$. 
The equations solved are
$$U^\prime = U + A_U + C_U\times(U-B) + D_U\times Z$$
$$B^\prime = B + A_B + C_B\times(B-V) + D_B\times Z$$
$$V^\prime = V + A_V + C_V\times(B-V) + D_V\times Z$$
$$I^\prime = I + A_I + C_I\times(V-I) + D_I\times Z,$$
where $A$ is the zero point in the respective filter, $C$ is the color term, 
$D$ is the extinction coefficient, and the primed magnitudes
are instrumental.
We list the best fit coefficients, observational scatter, and number of
stars used in the fits in Table 2. 
Data for each color come from two separate nights.
We find no significant variation between nights, so we combine
the data in order to best determine the zero point and color terms. 
The scatter introduced
by the calibration is fairly small, 1 to 3\%. 

\subsection{Internally Estimated Uncertainties}

\subsubsection{Photometry}

To determine the stability of the DAOPHOT photometry, 
we compare the photometry for a set of stars in the
overlap regions between scans or subscans.
One concern is that the photometry may vary
between scans or within a scan due to PSF variations. 
The best available internal test comes
from the comparison of adjacent scans. Adjacent scans are typically
from different nights and so have different seeing, extinction,
and focus, as well as completely independent reduction. 

The internal photometric stability can be demonstrated in at least
two ways. First, 
the absence of any obvious discontinuities in the magnitude
differences for stars observed in scan overlap regions,
plotted in Figure 2, argues that the reduction of subscan
images is stable. There is no evidence in Figure 2 that photometry
from different subscans varies significantly.
Of all the scans, only scan 70 in $I$ appears to 
have serious photometric problems, but those are most likely the
result of nonphotometric conditions. Second, the magnitude
differences between matched stars are in rough agreement with the
internally estimated uncertainties. In Figure 3, we present the
uncertainty
normalized differences between magnitudes in the two scans after
the polynomial photometric offset is applied. If the
propagated uncertainties are a true and complete reflection of the
uncertainties (excluding uncertainties in the photometric
calibration), these histograms should approach a Gaussian with
a dispersion of one. We plot such Gaussian curves 
for comparison. In most cases, the propagated
uncertainties only moderately 
underestimate (0 to 40\%) the true internal uncertainties.
The values of the quantity
$${\chi^2\over N-D} = 
{1\over{N-D}}\sum_{i}^N {(m_1 - m_2)^2_i\over \sigma^2_{m_1-m_2,i}},$$
are given in Table 3, where $D$ represents the degrees of freedom in
the fitted polynomial, $N$
the number of stellar pairs, $m_1$ and $m_2$ the magnitudes for the given
star in each of the two scans, and $\sigma$ the propagated
uncertainty in the magnitude error.
Including scan 70, the values of $\chi^2/(N-D)$ suggest
that the uncertainties are underestimated
by a factor $\le$ 2, and are typically underestimated by
less than a factor of 1.5.

We conclude that 
scientific results that are robust to a factor of two change in the current
uncertainties are trustworthy. Eventually, with additional scans,
we will be able to ascertain whether the true uncertainty is up to a factor
of two larger than estimated by propagation of DAOPHOT error
estimates, or whether a subset of the scans (\eg scan 70)
should be observed again.

\subsubsection{Astrometry}

We estimate the astrometric precision by examining the
distribution
of {\it rms} residuals between our coordinates 
and the cataloged positions of the MACS stars among the subscans.
In Figure 4, we plot a histogram of the
residuals in the astrometric solutions of all of our
72 subscans. The mean {\it rms} scatter
is below half an arcsec. The positional uncertainty
of the MACS stars is moderate (generally $\ltsim$ 0\sec5; Tucholke, de
Boer, \& Seitter).
Given our range of {\it rms} residuals, 
we do not appear to be introducing significant
uncertainties from the application of our
gnomic-projection coordinate transformation. 

The matching of subscans and scans, which uses data from
the edges of subscans and scans,
places the strictest demands on the coordinate solutions.
If stars at the edge of subscans have significantly larger
astrometric
errors, then fewer stars will be matched along subscan edges than in
the middle.
Such a trend is not observed (see below). 
On the other hand, if the stars at the edges of entire scans
have significantly higher astrometric errors, then observations of the 
same stars in overlapping scans would not be matched and 
those stars would appear twice in the catalog. 
We find that if we adopt a search radius of 
3 times the {\it rms} scatter of the astrometric standards
(a typical radius of about 1\sec2)
for subscan or scan matching, 
we see no significant 
spatial variation that correlate with the edges of the subscans or
scans
in the number of stars in the final catalog for 
stars with $V \le 21$ (737,840 stars).
We do see
such variations for a choice of a matching radius of two times the
$rms$ positional difference, suggesting that some positions at the edges of 
the subscans may have errors $\sim$ 0\sec8. As we discuss below,
the completeness of the survey drops sharply at $V \sim 21$ and since
different scans, and even different subscans, can have different
magnitude limits due to changes in the seeing and stellar density, 
there will be apparent discontinuous stellar
density differences for stars with $V > 21$,
even if the astrometric positions are accurate.

To qualitatively demonstrate the spatial uniformity of the final catalog,
we produce a stellar density image of the catalog. 
We include all stars with $V \le 21$, and bin the density image
into 15 arcsec square pixels. This image is 
then smoothed with a Gaussian function with $\sigma =$ 2 pixels to bring out
low signal features in the distribution. The use of the stellar
density weights this image toward stars near the magnitude
cutoff and so is most sensitive to spurious detections and
mismatches. This stellar density image is shown
in Figure 5. The union between subscans and between scans
is nearly seamless (the principal blemish, the cross feature just
north of the dominant stellar association in the field
is due to a super-saturated star that contaminated
the nearby photometry). 

\subsection{Externally Estimated Uncertainties}

The most challenging test of the data is a comparison to previously
published results by other investigators. Our direct comparison to 
published $UBV$ CCD photometry (Oey 1996) of the LH 38 region 
to a limit of $B \sim 20$ is presented
in Figure 6.
We matched stars between our list and Oey's list
using the same algorithm employed to match the photometry from
various scans
with fixed search radius of 2\sec0.
We match 252 of 274 stars in Oey's list in $B$ and $V$ and 103 of 111 in $U$. 
The stars that are unmatched are generally in crowded regions
and are either missing or identified at sufficiently different
positions.
Finally, we note that 
this region spans the edges of two of our scans, and so acts as a further
test of the scan-matching algorithms.

The agreement between Oey's photometric data and ours is generally
excellent,
despite the nebulosity and high stellar density in this region.
The zero point deviations, including all matched stars that are within
3$\sigma$ of the one-to-one correspondence line, are 0.064, $-$0.049, 
and $-$0.039 magnitudes for $U$, $B$, and $V$. Aside from the 
zero point deviations, there is a slight,
systematic
deviation at the bright end, especially in the $V$ filter (see lower
panel of Figure
6). The
deviation is of the sense that Oey's photometry underestimates
the magnitude relative to our measurement, 
possibly suggesting the onset of non-linearity in Oey's data.
This non-linearity is consistent with the appearance of the brightest
stars in Oey's
images which show hints of diffraction spikes. 
In her comparison to previous work
she obtained zero point magnitude offsets similar or larger than those
discussed above,
so that the offsets found between scan 62 and her data
are entirely consistent with those found among previous investigators.
Referencing our photometry to Oey's in the mean,
then
the $\chi^2$ values from the one-to-one correspondence line between
Oey's data and ours, excluding $> 5\sigma$ outliers, are 1.93, 1.50, and 
2.20, for $U$, $B$, and $V$ respectively.  These measurements of the
uncertainties
are only slightly larger than the results we obtained from our
comparison of overlapping scans. 
The histogram of 
uncertainty-normalized
deviations between our data and Oey's is shown in the upper panel of
Figure 6.

The results of this comparison
indicate that the photometry from the survey is reliable,
although our propagated uncertainties appear 
to underestimate the true external error by a factor of $\sim$ 1.5
(excluding zero point uncertainties and attributing the majority of the
$> 5\sigma$ outliers to incorrect matches). The random error is
the principal uncertainty of interest because
any zero point uncertainty will
be reduced with time as our survey area increases and comparisons with
more published photometry become possible.

\subsection{Super-Saturated Foreground Stars}

In a few instances, stars are sufficiently bright that their images
in the CCD scans
distort the photometry of objects in their vicinity. DAOPHOT 
identifies numerous spurious stars from 
fluctuations in the luminous halos of these stars.
For a star
to saturate so intensely it must have an apparent magnitude 
brighter than 10th magnitude. If such a star is in the LMC, then it 
would have an
absolute $V$ magnitude brighter than $-$8.5 ($M_{Bol} < -10.5$ for a
main sequence star), 
which is exceedingly rare (see Massey \etal 1995 for
examples). Therefore,
these heavily saturated stars are probably foreground stars. 
The photometry in the area surrounding 
these stars is most affected in the $I$ images.
We automate the search for these super-saturated
stars by making a $B-I$ color image of the stellar density using
15$\times$15 arcsec pixels. Because DAOPHOT
tends to identify many spurious stars that are bright in $I$ around
these problem stars, we search for regions of the density map that
are anomalously red (or that have large $B-I$). We first
median filter the image using a 2 by 2
pixel box to remove single luminous red stars.
We then run SExtractor
(Bertin \& Arnouts 1996) to identify bright sources (selected to 
have at least 10 pixels that are at least 2$\sigma$ above the mean), 
which we compare interactively
with the unfiltered image. We find 24 significant objects in the
entire
surveyed region.
These objects are typically annuli 
in the density image, where the
core is vacant because of saturation and the wings are densely
populated due to spurious sources.

We mask these regions using the results from SExtractor and
an interactive
determination of the mask size that we describe next. The mask
aperture should be related to the ``luminosity'' of the sum of spurious
stars, so we use the SExtractor ``magnitudes'' to set a scale for the
masks. We set the normalization interactively by considering a range
of normalizations and examining the color-magnitude diagrams of stars
identified within those mask apertures.
In Figure 7, we plot those CMD diagrams, beginning with the core
region (typically about 7 arcsec to 20 arcsec radius, depending on the
SExtractor ``luminosity''). Three larger annuli,
each of which contains roughly the
same number of stars as the inner annulus, are compared to the inner
annulus. The effect of the saturated stars
is especially evident in
the inner region, even among $B$ and $V$ photometry
where the effect is smaller than
in $I$. The upper main sequence and RGB clump tighten
substantially for annuli far from the saturated stars.
Conservatively, we will exclude a region around these saturated
stars equal to the area covered by all of the
panels in Figure 7, even though the last two panels appear to be
undisturbed by the saturated star.
The adopted mask apertures correspond to between 15 and 45 arcsec
in radius. This cut excludes only a small
fraction
of the total number of $V \le 21$ stars, 3,753 out of 737,840.

\subsection{Completeness}

As is always the case in crowded-field photometry,
some stars brighter than the detection limit are missed due to blending. We
use the standard artificial star algorithm from DAOPHOT to add stars
to the images at random and to estimate the
completeness of our sample by attempting to recover these artificial
stars. Five hundred stars over the full range of magnitudes
are added at random to two subscans from each
scan. We conduct five realizations, each with 500 artificial stars.
Instrumental magnitudes are remeasured for all of the stars in
the frame using the three-iteration DAOPHOT algorithm described
previously. By determining which artificial stars are recovered, we 
estimate our completeness as a function of magnitude in each filter.
In addition to identifying artificial stars as lost if no match is found,
we also identify stars as lost if the magnitude of the matched star
differs by more than 0.5 mag {\it and} if
such a discrepancy is significant at the greater than 5$\sigma$ level.
The general procedure of placing artificial stars at random within an image
neglects any correlation in the stellar distribution,
which is roughly valid for the overall survey, 
but grossly incorrect for stars in stellar
clusters.

In Figure 8, we present the derived completeness fractions for two
scans
(58, which has the highest mean stellar density, and 66, which has
the lowest stellar density among those observed in photometric conditions)
as a function of magnitude.
The Figure illustrates substantial
incompleteness ($> 50\%$) occurs only for magnitudes fainter than
21 in $B$ and $V$. In the Figure, we also include the magnitude at which the
completeness fraction drops below 0.5. 
Among the mean stellar densities covered in our present survey, there
are no large differences in completeness, 
with the possible
exception of the $I$ band data. The slight differences between scans are 
a result of various effects, including different stellar densities, seeing,
and a variable PSF. In general, we conclude that the survey is 
complete at the $>$50\% level for $B$,$V < 21$.

Despite this general conclusion regarding completeness,
a visual inspection of the stellar density image clearly shows that
the incompleteness can be rather severe even 
for relative bright stars in the densest regions. 
For example, there is a clear absence of red clump
stars ($V \sim 19$) at the center of the richest clusters in the
region surveyed. Therefore, completeness corrections should be
calculated in detail for any analysis of the survey data that requires
a complete stellar sample or detailed knowledge of the selection effects.

\section{Discussion}

The data obtained from this survey will be used to address 
a wide range of issues ---
from the study of the spatial variations in star formation history
to identifying background quasars for study of the low density
ISM. We are currently studying the extinction in this region (Harris,
Zaritsky, \& Thompson 1997) and the spatial distribution of
stars.

The Hess diagram for this region of the LMC is shown in Figure 9.
We construct this diagram by modeling each star with a Gaussian
of width along each axis corresponding 
to its observational uncertainty along the
axis. We sum the Gaussians and take
the square root of the intensity in each pixel
to enhance the contrast of the diagram in Figure
9. Data for over 1 million stars identified in both $B$ and $V$ 
are incorporated in this Figure. A complex star formation history
is evident: both a significant upper main sequence
and a red giant branch are visible. In the future, we will
present a detailed analysis of the constraints imposed by these data
on the star formation history.

\subsection{LMC Clusters}

One straightforward use of these data is to identify star
clusters. Given the digital nature of the data, we can easily construct
density maps that are independent of nebular emission and that utilize
color and magnitude information. We have previously discussed the 
stellar density image shown in Figure 5. Each pixel is 15 arcsec
on a side and the intensity of each pixel reflects the 
{\it number} of stars in that pixel that are in the catalog (\ie
stars identified in both $B$ and $V$).
This map of the stellar density
enables us to select clusters simply on the basis of stellar
density rather than being biased by the presence of luminous stars.
We only use stars with $V < 21$ mag and 
apply no color criteria to generate Figure 5.
We proceed by identifying significant
concentrations of stars. We create one version of the density 
image that is smoothed
using a Gaussian with 2.5 pixel width and another image smoothed
with a Gaussian 3 times as wide (7.5 pixels). We then subtract the latter
from the former to remove the smooth background and accentuate the
clusters (\ie unsharp masking). The image analysis package SExtractor
(Bertin and Arnouts 1996) is used to identify significant objects in 
this image. Our objective is to maximize the number of
objects detected, while minimizing spurious detections.
For detection, we require than an object contain at least
5 significant pixels and
we examine three choices for the threshold at which a pixel is 
considered significantly above the mean (3, 4, and
5$\sigma$). By applying our three choices of significance
thresholds to the image, we identify 157, 78, and 45 objects,
respectively.
To determine the number of spurious detections in this procedure,
we invert the image and search for positive detections. In the
negative image, we
find 31, 6, and 0 detections. Of the six detections found with the
choice of a 4$\sigma$ significance threshold, 
five were adjacent to the two most significant
clusters NGC 1846 and NGC 1851 and result from the dark halos generated by
the unsharp masking process.
Therefore, we conclude that the choice of 4$\sigma$ maximizes the
number of detections, while keeping the number of false detections to 
fewer than one or two.

In Figure 10, we present the unsharp mask image used to identify clusters
and our cross identification with objects in the Kontizas \etal 
atlas (1990).
Objects marked SP are suspected to be spurious because
they lie at the boundaries between scans or subscans, and because 
the density at those boundary regions can be nonuniform due to 
slight matching difficulties (the unsharp masking technique 
accentuates these boundaries). The objects marked AN (for anonymous)
are identified
as new clusters, for which no identified counterpart was available in
the Kontizas \etal catalog.
Objects marked Ce are clusters found in what was previously identified
as an emission line region in the Hodge and Wright atlas (1961; from identifications
by Henize 1956). Objects marked K are from the Kontizas \etal catalog.
Within this region, we identify 38 of the 44 clusters in the
Hodge and Wright
atlas, 9 additional clusters found by Kontizas \etal,
and 21 new clusters. Note that some of the clusters (eg. 
K 544 and K 622 were identified by Hodge (1988) as 131 and 188
respectively) and that AN 4 is close to, but not exactly at,
the position of Hodge's (1988) cluster 118. Table 4 lists
the positions from our astrometry,
the previous identification if available from the
Kontizas \etal catalog, and
a measure of the total stellar content of the clusters, $m_D$,
($\equiv -2.5$log(stellar content)) for intercomparison between
clusters.
The richness measure, $m_D$, is highly uncertain and biased low at the 
rich end due to crowding effects and saturation. 

Why do we miss certain clusters (NGC 1873, HS 130, HS 208, HS 222, HS 225, and
SL 289)? Some clusters, such as HS 130,
are visible in the unsharp masked image, but they are
insufficiently significant to be detected with our particular chosen
detection
parameters. We test whether the cluster contrast is higher for
these particular clusters if either a brighter magnitude or bluer color
criterion is imposed. If we use $V< 19$ to generate 
the stellar density image, then we detect NGC 1873, 
HS 130, HS 208, and SL 289 from the list of previously missed clusters,
but we lose HS 154, NGC 1911, NGC 1915, SL 263, and SL 269 as well 
as AN 2, 4, 9, 10, 11, 13, 15, and 17, presumably because these
clusters have proportionally fewer luminous stars. 
If we set our criteria to include upper main sequence stars and
exclude evolved stars,
$V>20$ and $-0.4 < B-V < 0.6$, then we identify
NGC 1873, HS 130, HS 208, and SL 289, but we lose NGC 1895 and
AN 2, 5, 11, and 15. Finally, selecting clusters only on the basis of
the density distribution of red clump stars, we find only 14 significant
objects, two of which are on the very edge of the image and clearly 
unreliable. Therefore, in contrast to the 68 clusters identified 
using all of the stars, there 
are only 12 statistically significant concentrations of red clump
stars. The detections correspond to NGC 1806, SL 174, AN 5, SL 298,
NGC 1871, NGC 1852, NGC 1829, NGC 1846 (3 
of the additional objects are unmatched objects and 1 is
spurious).
This list includes some of the richest clusters (\eg NGC 1806,
NGC 1846, 
SL 298, and NGC 1852), and so we suspect that these are detected
only because they are so rich to begin with that a large concentration
of red clump stars is found.
The three clusters that are only detected on the basis of their red
clump
star concentrations may be more representative of the field stellar population.
We will discuss the use of CMD diagrams to 
help further define the 
cluster star formation history of the Clouds 
elsewhere.

\section{Summary}

We discuss the first data from a, 4-band, CCD survey
of the central 8\degrees\ $\times$ 8\degrees\ of the LMC and 4\degrees\
$\times$ 4\degrees\ 
of the SMC. We have developed an automated data reduction pipeline 
that utilizes DAOPHOT II  to reduce drift scan images obtained using
the Great Circle Camera at the Las Campanas 1-m Swope telescope. 
The photometric uncertainties, based on both internal
and external comparisons of stellar magnitudes, 
are at most a factor of two worse than estimated from
the internal, propagated uncertainties generated by DAOPHOT and the
associated reduction pipeline.
Although a comparison with published photometry
can only currently be done for one region, LH 38 (Oey 1996), the
photometry is consistent among the two studies.
As our survey grows, there will be many overlapping regions available
both to confirm the automated reduction and to accurately fix the 
photometric zero point.
Astrometric precision appears to be roughly 0\sec5 for most of
the survey. Although 
astrometric and photometric accuracy is dependent on seeing,
there are no large-scale variations in data quality over
the observed area (with the exception of one scan taken in 
non-photometric conditions). Completeness is greater than 50\%
for $V \ltsim 21$ and we identify some stars 
down to $V \sim 22$. The other bands are observed to comparable limits.

The data are not yet publically available because we expect that
some slight modifications to the data reduction pipeline may be 
necessary after more data, especially in regions of 
higher stellar density, are obtained. Those interested in
the data described here should contact the first author.

We search the data for previously undiscovered star
clusters
by examining the stellar density image of the region. 
We use unsharp masking techniques on various subsamples of the data
to construct a cluster catalog that increases the number of 
identified clusters by about 45\%. The use of the stellar density,
particularly at magnitudes of $V \sim 20$, allows us to identify clusters
that are not easily identified solely on the basis of the upper main
sequence. Further study of these fainter clusters may help
clarify the cluster star formation history in the LMC. We do not
find a significant population of clusters with a large red clump
stellar population (only three, in comparison to 68 clusters, might lack an
upper main sequence and have a strong red clump population).
This findings are consistent with a 
lack of intermediate age-relative metal rich
clusters 
and the large numbers of young clusters
(age $\le$ 4 Gyr; \eg Olszewski, Suntzeff, Mateo 1996 and references therein). Additional
inferences await the detailed analysis of the cluster CMDs.

This paper describes the first steps toward a complete,
digital, photometric, $UBVI$ survey of the LMC and SMC. The quality of the
survey is such that a variety of questions can be addressed with
the data, ranging from the identification of star clusters, as done here,
to more complicated issues regarding the structure and evolution
of the Magellanic Clouds. In combination with an ongoing emission
line survey (Smith \etal 1996) 
and other surveys of the Clouds,
these data should provide new insight into our nearest galactic
neighbors.

\vskip 0.5in

\noindent

Acknowledgments: DZ gratefully acknowledges financial support from
a NASA LTSA grant (NAG-5-3501) and an NSF grant (AST-9619576), 
support for the construction of the GCC
from the Dudley Observatory through a Fullam award and a seed grant
from the Univ. of California for support during
the inception of this project. We thank A. Zabludoff for a comments
on a preliminary draft. We also thank
C. Anderson and S. Gibson for making available the image in Figure 1, 
originally obtained by Henize.
DZ also thanks the Carnegie Institution
for providing telescope access, shop time, and other 
support for this project, and
the staff of the Las Campanas observatory, in particular Oscar Duhalde,
for their usual excellent assistance.

\clearpage

\centerline{References}

\apj{Alcock, C. \etal 1996}{461}{84}
\aa{Bertin, E., \& Arnouts, S. 1996}{117}{393}
\annrev{Feast, M., \& Walker, A.R. 1987}{25}{345}
\mn{Gardiner, L.T., Hatzidimitriou, D. 1992}{257}{195}
\refbook{Harris, J., Zaritsky, D., \& Thompson, I. 1997, in prep.}
\mn{Hatzidimitriou, D., Hawkins, M.R.S., \& Gyldenkerne, K. 1989}{241}{645}
\apjsup{Henize, K.G. 1956}{2}{315}
\refbook{Hodge, P.W., \& Wright, F.W. 1967, ``The Large Magellanic
Cloud'',
(Smithsonian Press: Washington, D.C.)}
\pasp{Hodge, P.W. 1988}{100}{1051}
\refbook{Kontizas, M., Morgan, D.H., Hatzidimitriou, D., \& Kontizas, E
1990, AA Sup., 84, 527}
\aj{Landolt, A.U. 1992}{104}{340}
\aa{Lequeux, J., Rayo, J.F., Serrano, A., Peimbert, M., \&
Torres-Peimbert,
S. 1979}{80}{155}
\apj{Massey, P., Lang, C.C., DeGioia-Eastwood, K., \& Garmany,
C.D. 1995}{438}{188}
\aas{Oestreicher, M.O., \& Schmidt-Kaler, T. 1996}{117}{303}
\apjsup{Oey, M.S. 1996}{104}{710}
\annrev{Olszewski, E.W., Suntzeff, N.B., \& Mateo, M. 1996}{34}{511}
\aa{Schwering, P.B.W.,  \& Israel, F.P. 1991}{246}{231}
\refbook{Smith, R.C., \etal 1996, BAAS, 188, 5101}
\pasp{Stetson, P.B. 1987}{99}{191}
\aa{Tucholke, H.-J., de Boer, K.S., \& Seitter, W.C. 1996}{119}{91}
\pasp{Zaritsky, D., Shectman, S.A., \& Bredthauer, G. 1994}{108}{104}

\clearpage

\newpage
\centerline{\bf Figure Captions}

\noindent
\figcaption{unavailable electronically}
{The area of the LMC observed in this work is shown
as a box overlayed on an greyscale image of the LMC, with each scan
numbered. North at the top, East to the left. The rectangular box
subtends approximately 2\degrees\ by 1.5\degrees. }

\medskip
\noindent
\figcaption{}{The photometric matching of adjacent scans using 
stars common to both scans. The scan pair is labeled
at the top of each column, and each column contains the comparison
for each of the four filters.
The solid line
illustrates the polynomial fit used to register the photometry. Note that
the scale is vertically shifted for the $I$ matching of scans 66-70
(labeled to the right of the plot). \label{Figure 2}}

\medskip
\noindent
\figcaption{}{The uncertainty-normalized magnitude differences for
stars matched between different scans. The scan pair is labeled at
the top of each column, and each column contains the comparison
for each of the four filters. The solid curves are Gaussians with
unit dispersion.\label{Figure 3}}

\medskip
\noindent
\figcaption{}{The distribution of astrometric $rms$ differences between
our gnomic-projection equatorial coordinates and those in the 
Magellanic Catalogue of Stars (Tucholke, de Boer, \& Seitter 1996)
catalog for all 72 scans.\label{Figure 4}}

\medskip
\noindent
\figcaption{available at http://www.ucolick.org/instruct/lmcdir/fig5.jpg}{A stellar density plot (1 pixel = 15 arcsec)
for all stars with $V \le 21$. Corresponds the region of the four
scans shown in Figure 1. North at the top, East to the left.
The image is roughly 2\degrees\ across.}

\medskip
\noindent
\figcaption{}{Comparison between photometric data from Oey (1996) and
the survey data. Lower panels contain differences in magnitudes as a
function of magnitude. 
Solid lines indicate mean offsets in photometry.
The panels are $U$, $B$, and $V$ data left to right.
Upper panels show the distribution of magnitude errors normalized
by the propagated internal uncertainty. The solid curve
represents a Gaussian with $\sigma = 1$. In the $V$ panel, the thinner
line represents a Gaussian with $\sigma = 2$.}

\medskip
\noindent
\figcaption{}{The color-magnitude diagrams for regions around supersaturated
stars. Each panel contains roughly the same number of stars
in annuli of increasing radii. The effect on the photometry from
the saturated stars is gross in the innermost section and negligible
in the last two panels. We mask a
region
equivalent to the area encompassed by 
all four sections from further analysis.}

\medskip
\noindent
\figcaption{}{Results from artificial star experiments. The completion 
fraction as a function of magnitude is shown for subscans of
scans 58 and 66 for the four filter bandpasses. Included in each
panel is the magnitude at which the completeness fraction drops below
0.5.}

\medskip
\noindent
\figcaption{unavailable electronically}{A Hess diagram (a density-weighted CMD) for $B$ and
$V$. The left panel shows the greyscale version of 
the square root of the stellar density (for the purpose of reducing
contrast) across the color-magnitude plane, 
while the right shows a surface plot of the true density distribution
to illustrate the relative number of stars in the various
components.\label{}}

\medskip
\noindent
\figcaption{unavailable electronically}
{An unsharp-masked image of the stellar distribution
shown in Figure 5, with
clusters labeled. Same region and orientation as in Figure 5.
Objects labels are described in the text.}

\noindent
\clearpage

\begin{table}
\caption{Scan Coordinates}
\halign{\hfil#&\qquad\qquad\hfil#&\hfil#&\hfil#&\qquad\hfil#&\hfil#&\hfil#\cr
\noalign{\bigskip}
\noalign{\hrule \vskip 2pt \hrule \medskip}
\noalign{\noindent Scan \hskip 1.7cm $\alpha$ \hskip 0.2cm  (2000.0)
\hskip 0.5cm$\delta$}
\noalign{\medskip \hrule \medskip} 
58&05 &00 &28&$-$67 &53 &10\cr
62&05 &00 &46&$-$67 &31 &36\cr
66&05 &01 &02&$-$67 &9 &46\cr
70&05 &01 &19&$-$66 &47 &27\cr
\noalign{\medskip \hrule}
}

\end{table}
\clearpage
\begin{table}
\caption{Photometric Solutions}
\halign {#\hfil\quad&\hfil#&\quad\hfil#&\quad\hfil#&\quad\hfil#&\quad\hfil#\cr
\noalign{\bigskip}
\noalign{\hrule \vskip 2pt \hrule \medskip}
Filter&A&B&C&$\sigma_{CAL}$&$N_S$\cr
\noalign{\smallskip \hrule \medskip}
$U$&5.335&$-$0.122&$-$0.322&0.032&60\cr
$B$&3.557&$-$0.070&0.234&0.013&62\cr
$V$&3.271&0.059&0.174&0.011&67\cr
$I$&3.735&0.038&0.012&0.034&45\cr
\noalign{\medskip \hrule}
}

\end{table}
\clearpage
\begin{table}
\caption{Interscan Comparison}
\halign {#\hfil\quad&\hfil#&\quad\hfil#&\quad\hfil#&\quad\hfil#&\quad\hfil#\cr
\noalign{\bigskip}
\noalign{\hrule \vskip 2pt \hrule \medskip}
Scans&$\chi^2_U$&$\chi^2_B$&$\chi^2_V$&$\chi^2_I$&\cr
\noalign{\smallskip \hrule \medskip}
58-62&1.40&1.85&1.65&1.21\cr
62-66&1.36&1.90&2.38&1.00\cr
66-70&1.63&4.02&4.14&1.37\cr
\noalign{\medskip \hrule}
}

\end{table}
\clearpage
\begin{table}
\caption{Stellar Cluster Catalog}
\halign{#\hfil\quad&\hfil#&\hskip 5pt \hfil#&\hskip 5pt
\hfil#&\quad\hfil#&\hskip 5pt\hfil#&\hskip 5pt\hfil#&\quad\hfil
#&\qquad#\hfil\quad&\hfil#&\hskip 5pt\hfil#&\hskip
5pt\hfil#&\quad\hfil#&\hskip 5pt\hfil#&\hskip 5pt\hfil#&\quad\hfil
#\cr
\noalign{\hrule \vskip 2pt \hrule \medskip}
\noalign{\noindent Name \hskip 1.5cm $\alpha$ \hskip 0.3cm (2000.0)
\hskip 0.3cm $\delta$ \hskip 1.1cm
$m_D$ \hskip 0.6cm Name \hskip 1.4cm $\alpha$ \hskip 0.3cm (2000.0)
\hskip 0.4cm $\delta$ \hskip 1.1cm $m_D$}
\noalign{\medskip \hrule \medskip} 
HS 86&5&1&4.8&$-$67&54&58&$-$4.05&SL 281&5&10&34.0&$-$67&7&36&$-$3.85\cr
SL 174&5&1&12.3&$-$67&49&7&$-$4.90&Ce26&5&10&44.9&$-$67&4&58&$-$3.15\cr
K 438&5&1&14.8&$-$67&32&18&$-$3.61&K 622&5&10&53.1&$-$67&28&18&$-$3.85\cr
SL 173&5&1&22.2&$-$67&17&43&$-$4.27&HS 154&5&10&57.2&$-$67&37&28&$-$3.60\cr
AN 5&5&1&29.5&$-$67&37&52&$-$3.61&SL 293&5&11&9.6&$-$67&41&6&$-$4.10\cr
SL 179&5&1&44.8&$-$67&5&47&$-$4.11&AN 10&5&11&16.5&$-$67&55&57&$-$3.47\cr
K 4523&5&1&48.8&$-$67&28&25&$-$4.07&SL 298&5&11&30.9&$-$66&58&32&$-$5.51\cr
NGC 1806&5&2&11.7&$-$67&59&9&$-$6.16&SL 300&5&11&40.5&$-$67&33&58&$-$4.48\cr
AN 2&5&2&26.1&$-$66&36&45&$-$2.93&AN 15&5&12&11.3&$-$67&16&6&$-$4.24\cr
AN 7&5&2&27.9&$-$67&48&5&$-$3.57&SL 310&5&12&31.0&$-$67&17&29&$-$3.73\cr
SP&5&2&38.2&$-$67&53&54&$-$3.99&NGC 1864&5&12&40.5&$-$67&37&25&$-$4.38\cr
SP&5&3&21.2&$-$66&58&39&$-$3.91&AN 16&5&12&53.8&$-$67&15&36&$-$4.28\cr
SL 197&5&3&34.4&$-$67&37&29&$-$4.50&AN 13&5&13&0.7&$-$67&25&6&$-$4.02\cr
NGC 1816&5&3&52.2&$-$67&15&37&$-$4.38&AN 14&5&13&1.7&$-$67&19&29&$-$3.88\cr
AN 8&5&4&0.2&$-$67&52&59&$-$3.36&SP&5&13&33.4&$-$67&10&27&$-$3.30\cr
AN 9&5&4&19.6&$-$67&47&50&$-$4.47&NGC 1871&5&13&45.6&$-$67&26&18&$-$5.06\cr
AN 6&5&4&48.1&$-$67&42&9&$-$3.77&SP?&5&13&46.3&$-$67&44&0&$-$3.85\cr
NGC 1829&5&4&56.7&$-$68&3&39&$-$4.77&SP&5&13&51.8&$-$67&11&35&$-$4.50\cr
Ce21&5&4&59.6&$-$67&34&27&$-$4.54&NGC 1869&5&13&54.9&$-$67&22&38&$-$4.78\cr
Ce20&5&5&18.8&$-$66&55&4&$-$3.52&AN 1&5&14&0.4&$-$66&48&24&$-$4.09\cr
AN 4&5&5&19.7&$-$67&16&7&$-$3.54&K 680&5&14&40.7&$-$67&12&9&$-$3.85\cr
SP&5&5&22.0&$-$67&10&48&$-$3.74&HS 194&5&15&38.7&$-$66&41&34&$-$3.71\cr
AN 3&5&5&53.5&$-$66&44&11&$-$3.58&K 698&5&15&40.7&$-$67&50&43&$-$3.81\cr
HS 114&5&6&3.0&$-$68&1&33&$-$4.04&SL 347&5&16&25.8&$-$66&49&24&$-$3.76\cr
HS 116&5&6&12.6&$-$68&3&57&$-$3.61&AN 12&5&16&42.9&$-$67&48&1&$-$3.74\cr
SL 228&5&6&28.2&$-$66&54&22&$-$4.86&NGC 1895&5&16&51.2&$-$67&19&51&$-$3.96\cr
K 544&5&6&42.3&$-$67&50&25&$-$4.19&SP&5&17&8.5&$-$66&37&42&$-$4.85\cr
SL 239&5&7&9.5&$-$66&40&9&$-$3.69&NGC 1897&5&17&30.9&$-$67&27&00&$-$4.65\cr
NGC 1842&5&7&18.5&$-$67&16&19&$-$4.73&Ce34b&5&17&33.5&$-$66&44&0&$-$4.05\cr
NGC 1844&5&7&30.3&$-$67&19&28&$-$5.12&Ce34a&5&17&39.6&$-$66&42&13&$-$3.68\cr
NGC 1846&5&7&34.3&$-$67&27&36&$-$6.36&SP&5&17&45.6&$-$66&37&28&$-$3.98\cr
SL 263&5&7&41.5&$-$66&38&10&$-$3.39&NGC 1902&5&18&17.9&$-$66&37&40&$-$5.54\cr
HS 121&5&7&46.5&$-$67&51&36&$-$3.36&NGC 1905&5&18&22.9&$-$67&16&40&$-$4.40\cr
SL 263&5&8&26.4&$-$66&46&15&$-$4.35&K 758&5&19&11.3&$-$67&29&22&$-$3.52\cr
K 583&5&8&48.2&$-$67&43&39&$-$4.20&HS 224&5&19&16.2&$-$67&5&57&$-$4.07\cr
NGC 1852&5&9&23.7&$-$67&46&44&$-$5.91&NGC 1911&5&19&25.8&$-$66&40&50&$-$4.06\cr
SP&5&9&24.8&$-$66&48&45&$-$4.19&NGC 1915&5&19&27.4&$-$66&44&24&$-$3.81\cr
SL 269&5&9&35.6&$-$67&48&33&$-$3.76&K 771&5&19&42.5&$-$66&49&59&$-$3.75\cr
AN 11&5&10&2.4&$-$67&39&29&$-$3.56&SP&5&20&11.5&$-$67&55&49&$-$4.23\cr
\noalign{\medskip \hrule \medskip} 
}

\end{table}
\clearpage
\begin{figure}
\plotone{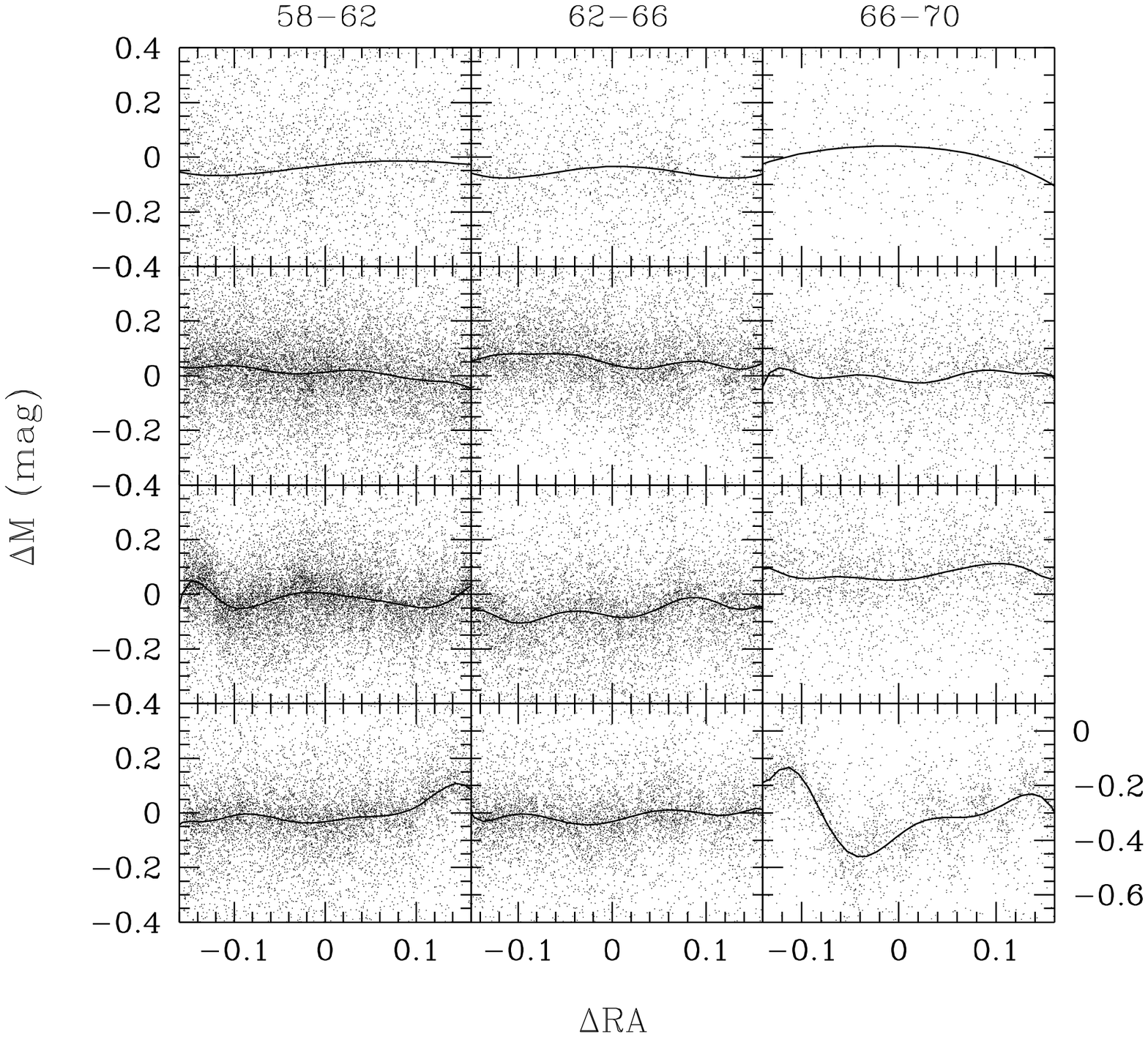}
\caption{Figure 2}
\end{figure}
\begin{figure}
\plotone{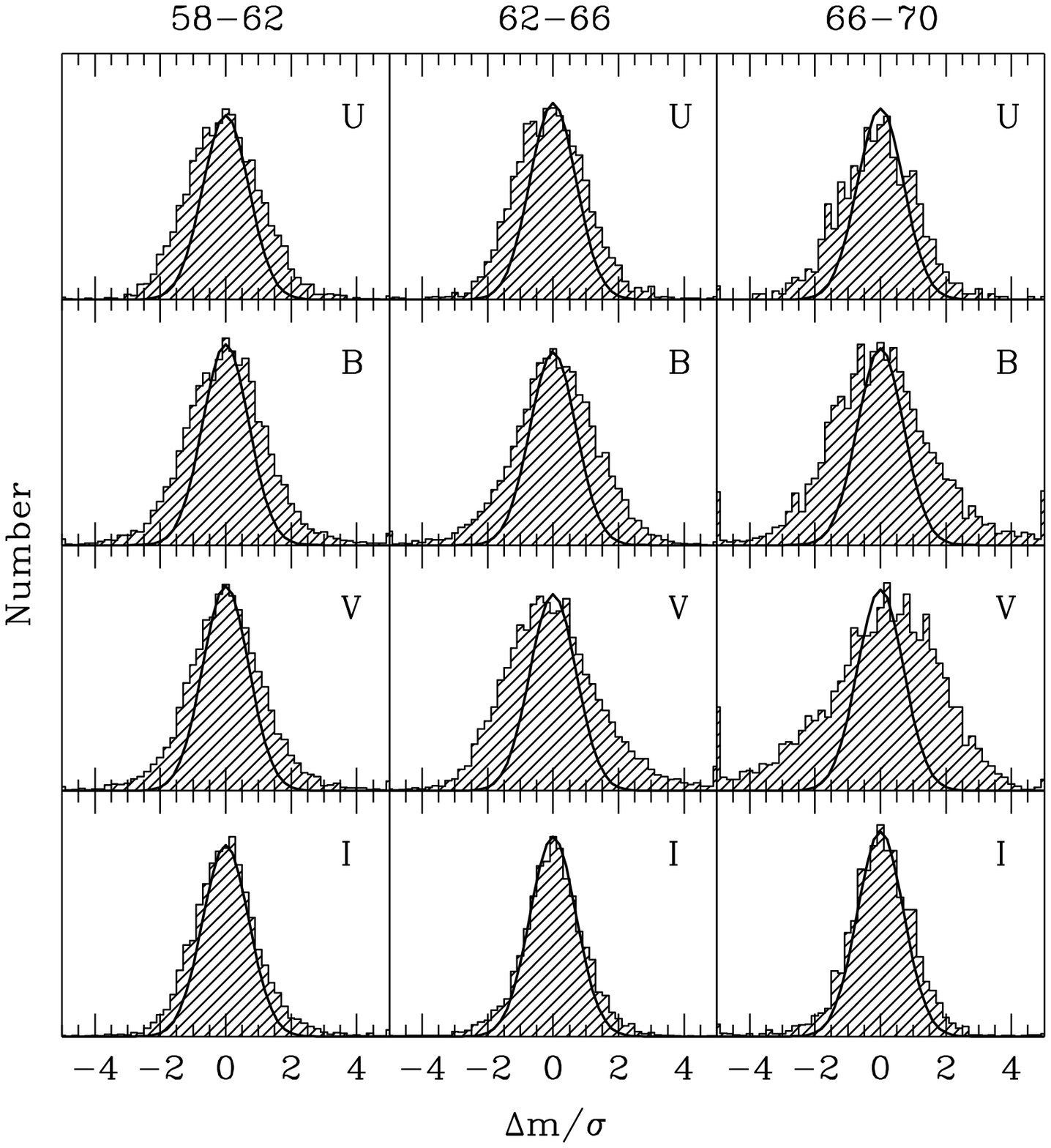}
\caption{Figure }
\end{figure}
\begin{figure}
\plotone{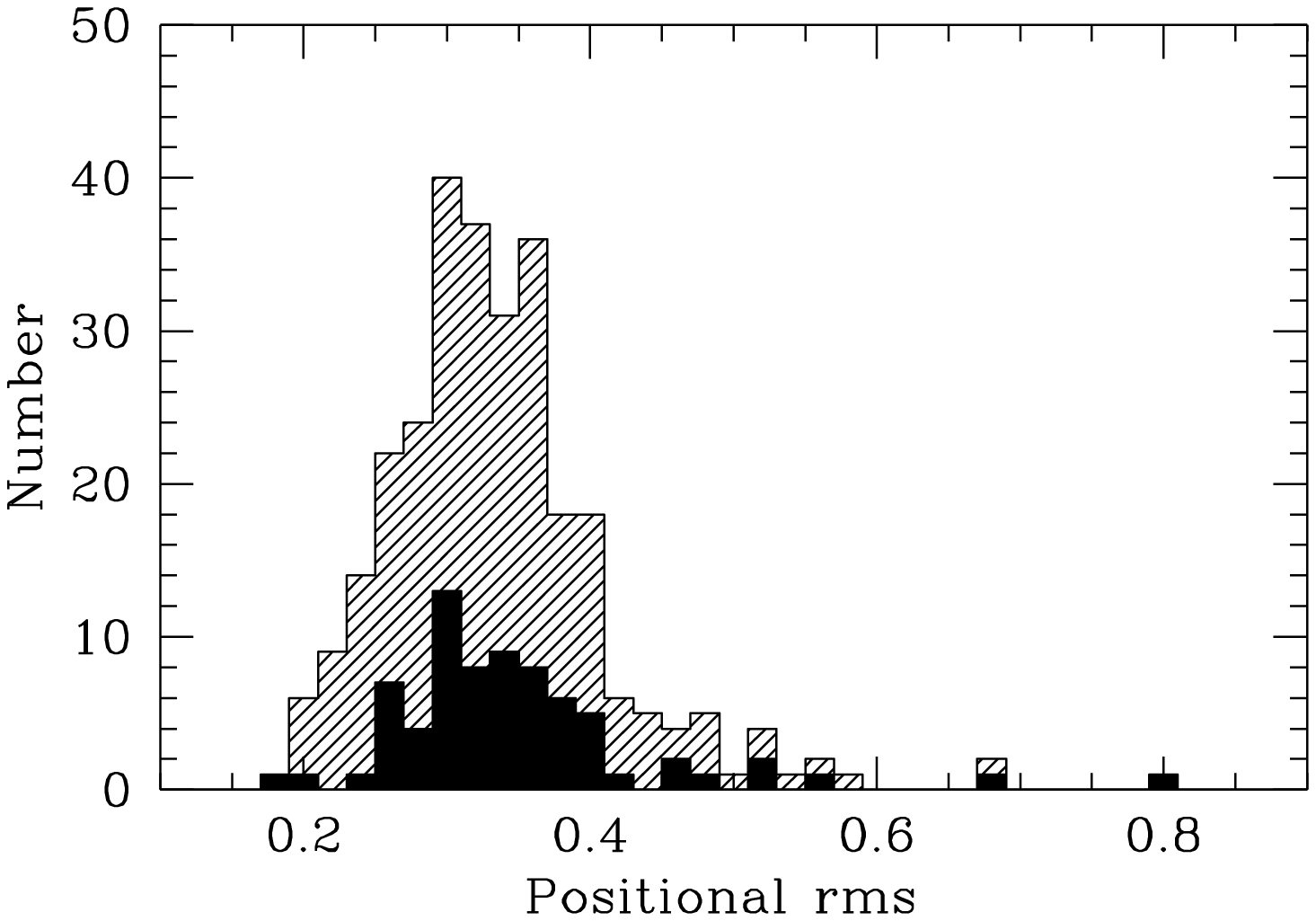}
\caption{}
\end{figure}
\begin{figure}
\plotone{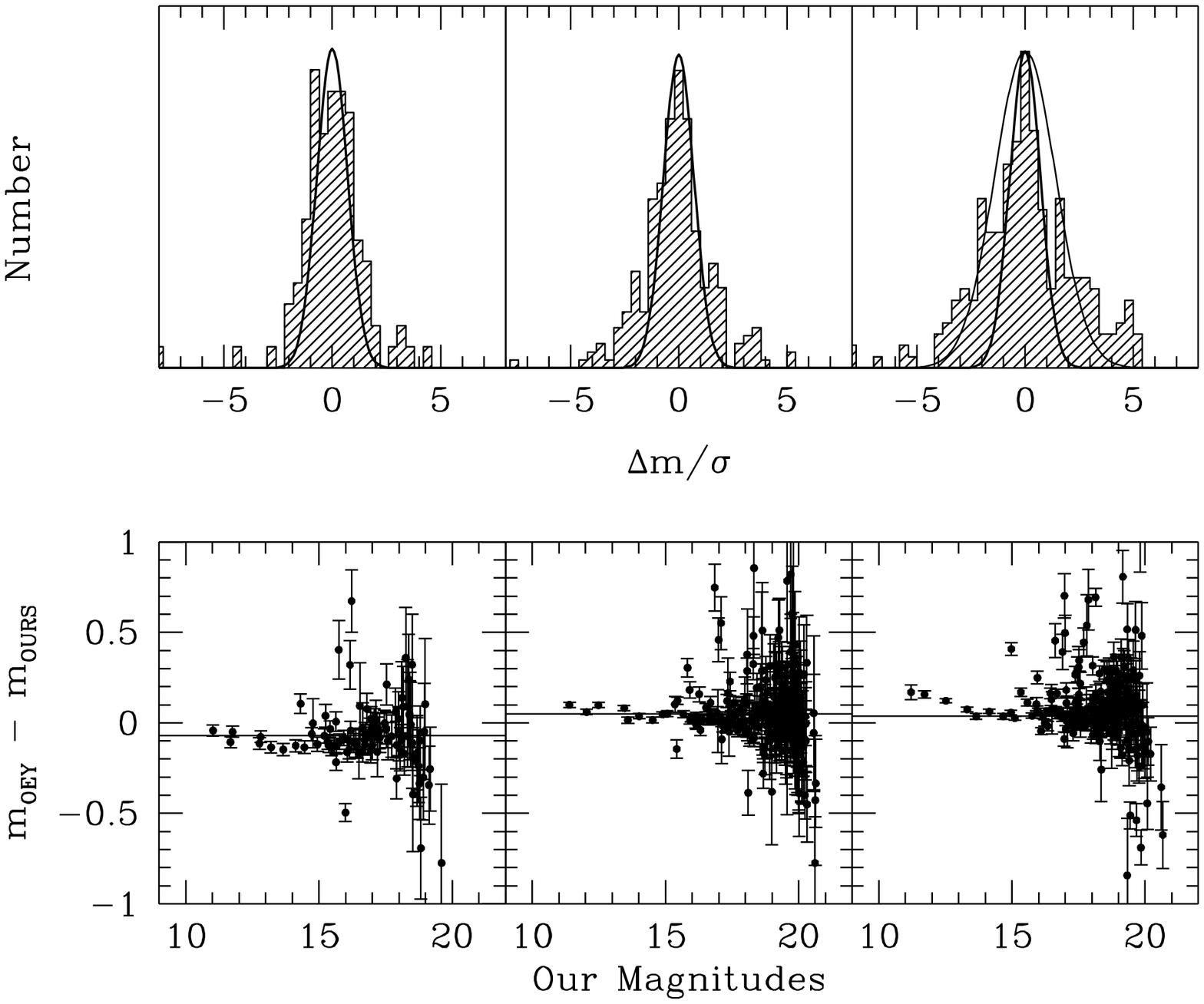}
\caption{}
\end{figure}
\begin{figure}
\plotone{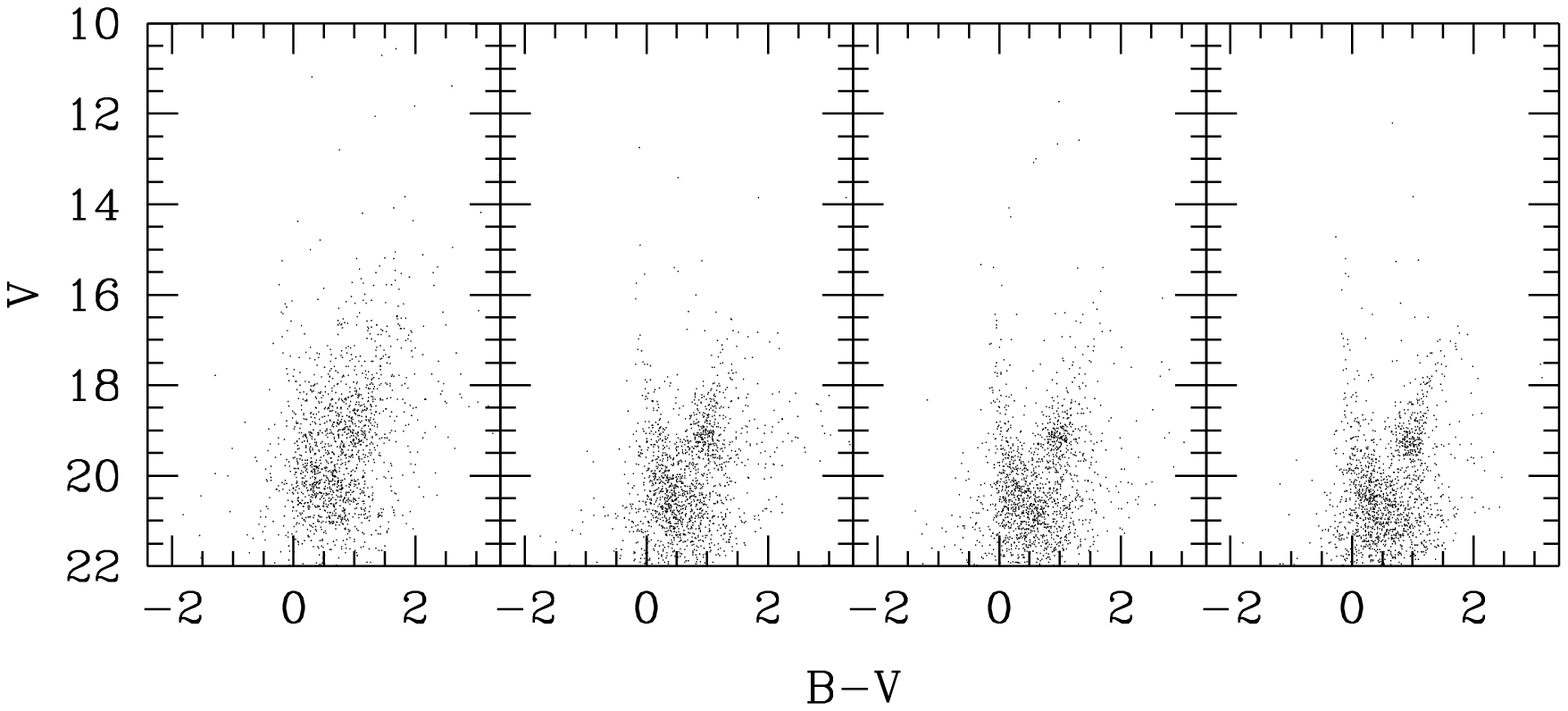}
\caption{}
\end{figure}
\begin{figure}
\plotone{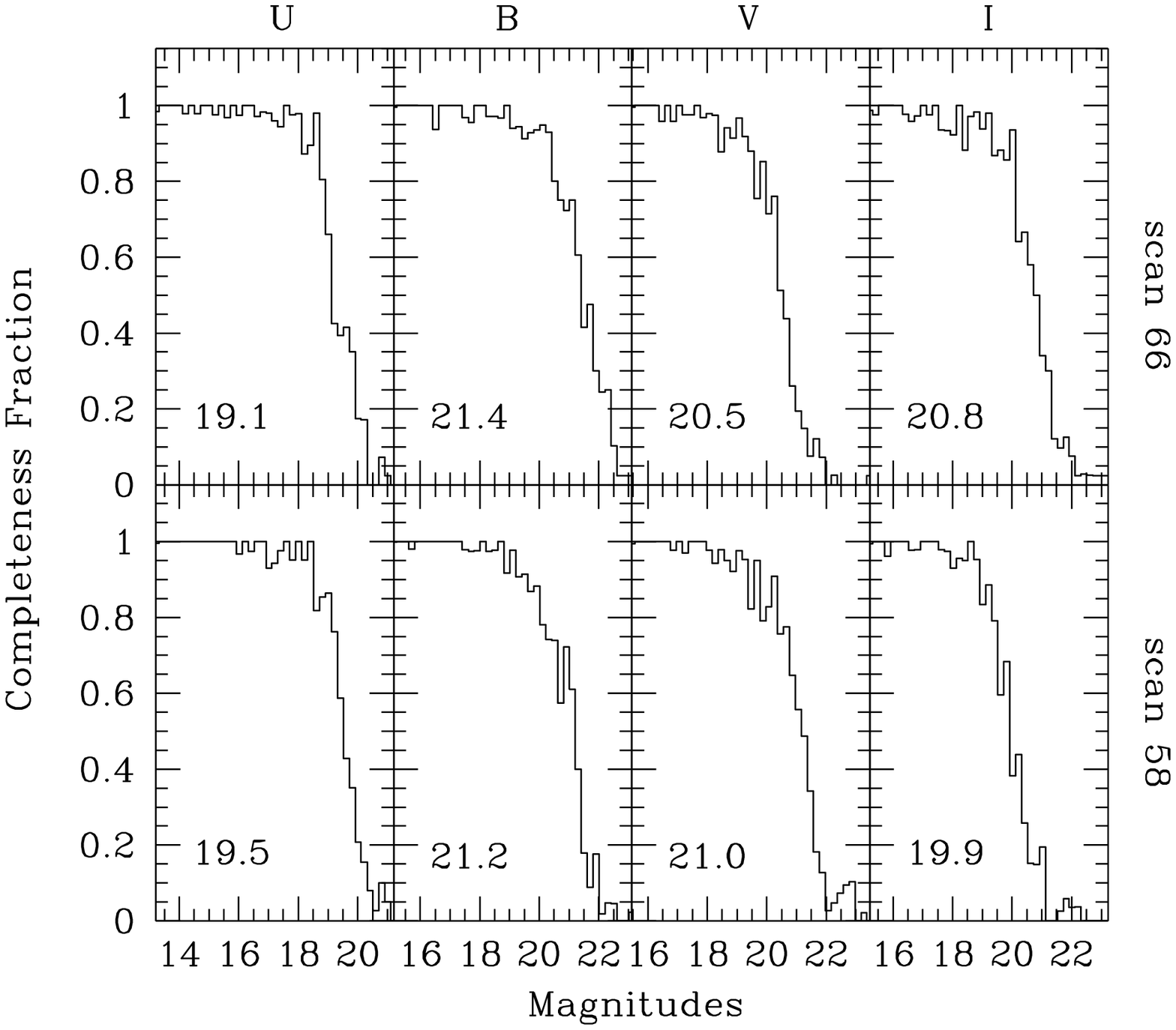}
\caption{}
\end{figure}
\end{document}